\documentclass{appolb}
\usepackage{epsfig}
\begin{document}
\title{Density of States for HP Lattice Proteins}
\author{Michael Bachmann\thanks{Email: michael.bachmann@itp.uni-leipzig.de}
\and
Wolfhard Janke\thanks{Email: wolfhard.janke@itp.uni-leipzig.de}
\address{Institut f\"ur Theoretische Physik, Universit\"at Leipzig, 
Augustusplatz 10/11, D-04109 Leipzig, Germany}
}
\maketitle
\begin{abstract}
The density of states contains all informations on energetic quantities
of a statistical system, such as the mean energy, free energy, entropy, and specific
heat. As a specific application, we consider in this work a simple lattice
model for heteropolymers that is widely used for studying statistical properties
of proteins. For short chains, we have derived
exact results from conformational enumeration,
while for longer ones we developed a multicanonical Monte Carlo variant
of the nPERM-based chain growth method in order to directly simulate 
the density of states.
For simplification, only two types of monomers
with respective hydrophobic (H) and polar (P) residues are regarded
and only the next-neighbour interaction between hydrophobic monomers, being
nonadjacent along the chain, is taken into account. This is known
as the HP model for the folding of lattice proteins.
\end{abstract}
\PACS{05.10.-a, 87.15.Aa, 87.15.Cc}

\section{Introduction}
Proteins perform numerous functions in a biological cell
system, e.g.\ controlling of transport processes of
organelles, stabilisation of the cell structure, enzymatic
catalysis of chemical reactions, etc. It is well established
that the three-dimensional conformation of a protein
within an aqueous environment determines its biological 
function. Due to the enormous number of tasks to be 
necessarily fulfilled to ensure the stability of a biological
system, a large number of various proteins exists. All of them
are built up of chains of amino acid residues, linked
by peptide bonds. Since $20$ different amino acids are known
from nature, a protein with $N$ monomers is, in principle, formed from
$20^N$ possible sequences. Only a small number of so-called
designing sequences, however,
is actually realised in equilibrium. The reason is that the protein must
be stable against thermal fluctuations and
may not fold into a different shape leading to a loss of its
associated function. Therefore, real proteins are supposed
to possess a funnel-like deep global minimum in a rough free energy 
landscape~\cite{dill1}. It is one of the essential goals of computational
protein research to identify the native state associated with
the global free energy minimum of a protein with a given sequence of amino acid 
residues.
Since the sequence of amino acids is known to be responsible for the
resulting fold, it is also interesting to analyse what
properties sequences of such favoured proteins have.   

Unfortunately, computer simulations of real proteins are
extremely difficult due to the relatively big number of 
degrees of freedom influenced by electrostatic, Lennard-Jones,
hydrogen bond, torsional, and environmental interactions
(for a review see, e.g., Ref.~\cite{okamoto}).
In order to qualitatively study the folding behaviour of 
proteins and also for sequence analysis, simple lattice
models seem to be very practical. Nevertheless, the
determination of the lowest-energy states and their
degeneracies remains challenging. 
In fact, it was shown~\cite{np_comp} that folding proteins
within the HP model~\cite{dill2}, the most simple lattice model for
proteins, is an NP--complete problem. 
On the numerical side, one technical problem
is that the polymers are required to be self-avoiding. Thus,
updating the conformation in a Monte Carlo simulation
is quite involved. 
Two completely different methods are 
widely used, first the application of a move set consisting
of transformations that allow the change of a conformation
of total length $N$, while in the second method, chain growth,
a new monomer is attached to the end of a partial chain
of length $n < N$ until the total chain length is reached.
Both techniques work well in computer simulations of
polymers at comparatively high temperatures, for example
the investigation of the $\Theta$-point transition between
compact globule polymer states and random coils~\cite{theta}. For studying 
the low-temperature behaviour of heteropolymers, however, 
the application of move sets is not very suitable, since 
transformations that usually 
belong to a move set, e.g.\ end and corner flips, crankshafts, 
and pivot rotations are inefficient for the creation
of very dense conformations. The transition between
lowest-energy states and compact globules represents a
``conformational barrier'' at low temperatures that is much 
better circumvented with chain-growth based algorithms
such as PERM~\cite{grass1} and its new variants 
nPERM$_{\rm ss}^{\rm is}$~\cite{hsu1}.  

We are interested in the energetic thermodynamic properties
of heteropolymers for all temperatures and therefore we 
proposed a multicanonical chain growth algorithm~\cite{bj1}
which allows an explicit sampling of the density of states.
The density of states is identical with the canonical
distribution at infinite temperature. 
Nevertheless, we also obtain very accurate results in the
low-temperature region which in effect is due to the
capacity of the multicanonical sampling~\cite{muca} which spreads 
the canonical distribution to a flat histogram, such
that all energetic states are, in principle, equally 
probable within the simulation. At the end,
the canonical distribution at any temperature and thus
all thermodynamic functions can be obtained by a simple
reweighting procedure. This is only possible,
since the multicanonical method allows a sampling of
the entire space of states, including such events that are
canonically suppressed by many orders of magnitude.
In our simulations of the HP model for lattice proteins with more 
than 40 monomers, we were also required to sample the 
lowest-energy states having a probability of realization
in the density of states of the order of $10^{-25}$, 
since these states dominate the
low-temperature behaviour of the protein. Another problem
is that the conformational transition between ground states 
and globules just appears in this temperature region, causing a
conformational barrier that is avoided best, as described above, 
by using an adequate chain growth algorithm. Therefore we combined
the multicanonical method with the new PERM variants for simple 
and importance sampling, nPERMss and nPERMis~\cite{hsu1}, respectively,
to obtain densities of states with high and uniform accuracies for all energies.
\section{Density of States of HP Lattice Proteins}
For simplicity, we investigate lattice proteins that consist of only two types
of monomers: hydrophobic (H) and polar (P). This choice is made
since most of the amino acids occurring in nature can be
grouped into these two classes. Moreover it is assumed that
the protein mainly folds due to an effective hydrophobic interaction.
This means that a core of hydrophobic monomers is formed which is 
screened from the aqueous solvent by a shell of polar (or hydrophilic)
residues. The simplest form of the HP model takes into account only
the attractive interaction between next-neighbouring H monomers
being nonadjacent along the chain~\cite{dill2}:
\begin{equation}
\label{hpmodel}
E = -\sum\limits_{\langle i,j<i-1\rangle} \sigma_i\sigma_j,
\end{equation}
where $\sigma_i = 0$ $(1)$ if the $i$th monomer is polar
(hydrophobic). The partition sum of a HP lattice protein
with fixed sequence at temperature $T$ is then given by 
$Z = \sum_{\{\bf x\}} \exp\{-E(\{{\bf x}\})/k_BT\}$, where the
sum is taken over all admissible conformations of the polymer.
Sorting all conformational states with respect to their
energies, the partition sum can also be expressed in terms of 
the density (or degeneracy) $g(E)$ of states with energy $E$:
\begin{equation}  
Z= \sum_i g(E_i)\exp\{-E_i/k_BT\}. 
\end{equation}
Knowing $g(E)$, the mean energy $\langle E \rangle$ of the system 
can be calculated by
\begin{equation}
\langle E \rangle(T) = \frac{\sum_i E_i\,g(E_i)\exp\{-E_i/k_BT\}}{\sum_i g(E_i)\exp\{-E_i/k_BT\}}
\end{equation}
and the specific heat is given by the fluctuation formula
\begin{equation}
C_V(T) = \frac{1}{k_BT^2}\left(\langle E^2\rangle-\langle E \rangle^2\right).
\end{equation}
Other energetic quantities being related to the density of states
are the Gibbs free energy
\begin{equation}
F(T) = -k_BT \ln \sum_i g(E_i)\exp\{-E_i/k_BT\}
\end{equation}
and the entropy
\begin{equation}
S(T) = \frac{1}{T}\left[\langle E \rangle(T)-F(T)\right].
\end{equation}
\section{Exact Enumeration of 14mers}
\begin{figure}[t]
\centerline{
\epsfxsize=4cm \epsfbox{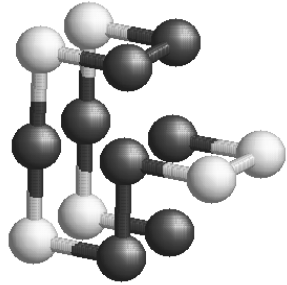}
}
\caption{\label{14.1} Unique ground state of the 14mer with designing sequence $14.1$ 
(dark spheres: hydrophobic residues, light spheres: polar monomers).}
\end{figure}
As a first example we have investigated HP proteins with 14 monomers
by enumerating all possible conformations.   
This study is quite interesting, because there is only one 
sequence (H\-P\-H\-P\-H$_2$\-P\-H\-P\-H$_2$\-P$_2$\-H, in the
following denoted as $14.1$) 
that is {\em designing}, i.e.\ the ground state of the
associated lattice protein is unique (up to translational,
rotational, and one reflection symmetry).  
\begin{figure}[t]
\centerline{
\epsfxsize=12cm \epsfbox{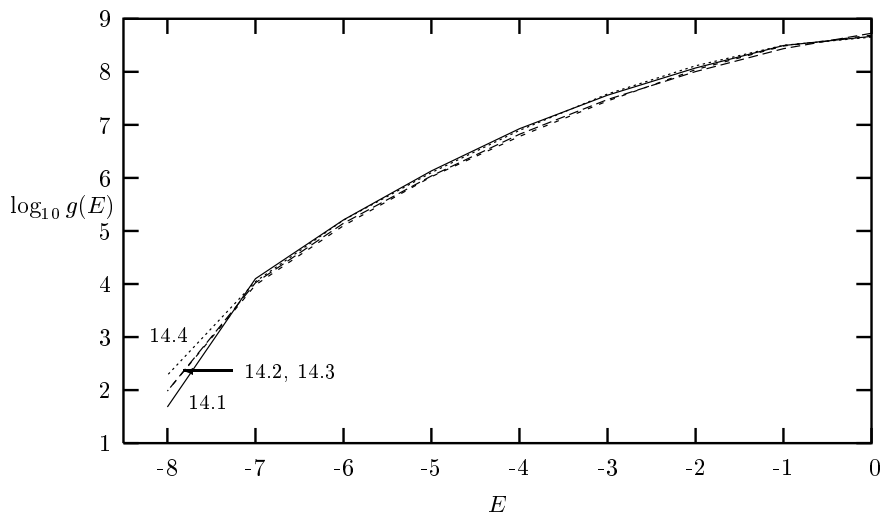}
}
\caption{\label{14dens} Exact densities of states of exemplified 14mers with
similar properties ($n_H=8$, $E_{\rm min}=-8$) but different sequence.}
\end{figure}
\begin{table}[b]
\caption{\label{T14} Exact total densities of the states with energy $E$ for the 14mers.
The entries of the table include all states that contribute to the partition function $Z_\infty$ for 14mers
at infinite temperature (except translations) which counts the number of self-avoiding
random walks with $(14-1)=13$ steps.}\vspace*{0.2cm}
\centerline{
\begin{tabular}{|r|rrrr|}\hline\hline
 & \multicolumn{4}{|c|}{sequence}\\ 
$E$ & 14.1 & 14.2 & 14.3 & 14.4  \\ \hline
$-8$ & 48 & 96 & 96 & 192\\
$-7$ & 12\,576 & 10\,560 & 9\,576 & 11\,136\\
$-6$ & 162\,120 & 140\,496 & 126\,240 & 160\,536\\
$-5$ & 1\,349\,808 & 1\,089\,792 & 1\,053\,744 & 1\,259\,040\\
$-4$ & 8\,434\,536 & 6\,661\,032 & 6\,028\,944 & 7\,831\,752\\
$-3$ & 36\,120\,840 & 29\,943\,792 & 28\,329\,504 & 38\,367\,360\\
$-2$ & 118\,052\,520 & 100\,663\,488 & 109\,433\,232 & 129\,351\,360\\
$-1$ & 312\,691\,992 & 273\,343\,176 & 305\,911\,056 & 314\,705\,352\\
$0$ & 467\,150\,070 & 532\,122\,078 & 493\,082\,118 & 452\,287\,782\\ \hline
$Z_\infty$  & 943\,974\,510 & 943\,974\,510 & 943\,974\,510 & 943\,974\,510 \\ \hline\hline
\end{tabular}
}
\end{table}
It possesses $n_H=8$ hydrophobic monomers and the ground-state energy is
$E_{\rm min}=-8$, since there are $8$ hydrophobic contacts (see Fig.~\ref{14.1}).
In order to understand the particular properties of such a protein
with the lowest ground-state degeneracy among all the $2^{14}$
different 14mers, we compare it with three other ones having
similar properties ($n_H=8$, $E_{\rm min}=-8$), but different
sequences and therefore different ground-state degeneracies. 
The degeneracy of the lowest-energy state of the sequences $14.2$ 
(H$_2$\-P$_2$\-H\-P\-H\-P\-H$_2$\-P\-H\-P\-H) and $14.3$ 
(H$_2$\-P\-H\-P\-H\-P$_2$\-H\-P\-H\-P\-H$_2$) is
twice that of the designing sequence $14.1$, while sequence $14.4$ 
(H$_2$\-P\-H\-P$_2$\-H\-P\-H\-P\-H$_2$\-P\-H) is even 
four times higher degenerated. 
\begin{figure}[t]
\centerline{
\epsfxsize=12cm \epsfbox{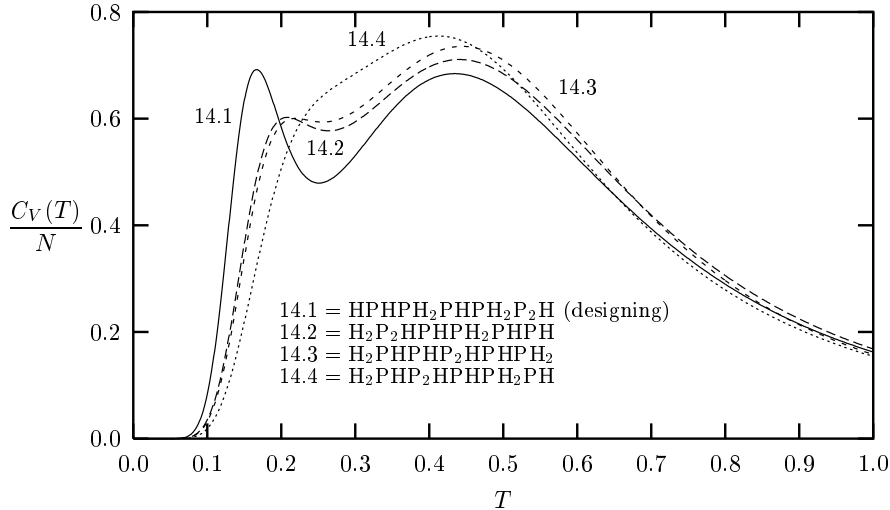}
}
\caption{\label{14C} Specific heat of the 14mers.}
\end{figure}
Figure~\ref{14dens} shows the 
densities of states for the four sequences. Since the densities of the excited
states do not considerably differ (see Table~\ref{T14}), the low-temperature behaviour of these proteins
can only vary due to the different ground-state degeneracies. Indeed, the specific
heat shown in Fig.~\ref{14C} exhibits a pronounced low-temperature peak only for the designing sequence $14.1$,
while it is largely suppressed for the other proteins. This peak indicates
the transition from the ground states to compact globule states. At higher temperatures,
the globules unfold and form random coil conformations.   
\section{Simulation of a 42mer: Lattice Model of {\em Pectate Lyase C}}
\begin{figure}[t]
\centerline{
\epsfxsize=12cm \epsfbox{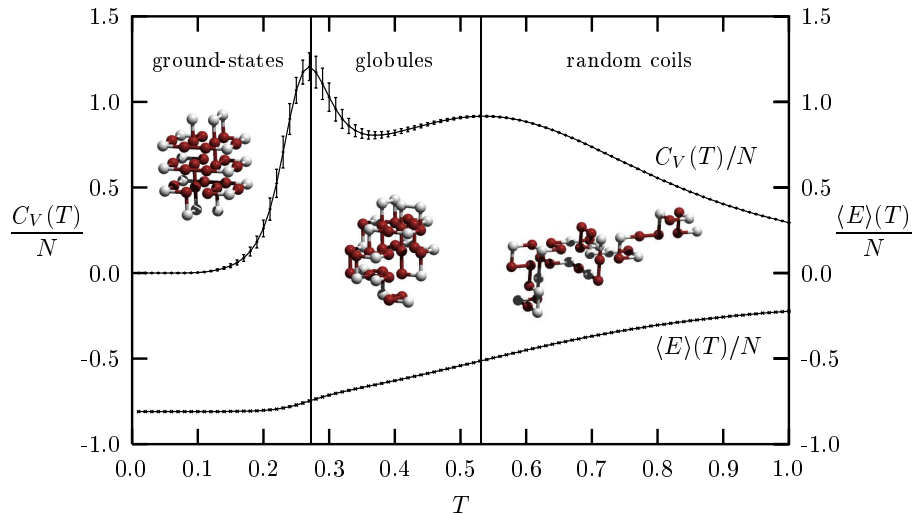}
}
\caption{\label{42C} Specific heat and mean energy of the 42mer.}
\end{figure}

For lattice proteins with more than 20 monomers, enumeration
becomes exhausting, since the number of conformations grows
exponentially with the number of monomers~\cite{irbaeck1}. More sophisticated 
search algorithms are required to sample the phase space.
For this reason, we developed a multicanonical chain growth algorithm~\cite{bj1}
that combines the advantages of avoiding conformational barriers
by using a PERM-based chain growth method~\cite{grass1,hsu1} and the capacity
of a flat histogram technique allowing the sampling of the 
entire energy space~\cite{muca}. In order to achieve this, the canonical
distributions provided by PERM at each intermediate length
of the growing chain must be flattened. As usual, the
multicanonical weights are determined by an iterative
procedure~\cite{muca}. 
We applied this method to calculate
the density of states of a lattice 42mer with sequence
PH$_2$\-P\-H\-P\-H$_2$\-P\-H\-P\-H\-P$_2$\-H$_3$\-P\-H\-P\-H$_2$%
\-P\-H\-P\-H$_3\-$P$_2$\-H\-P\-H\-P\-H$_2$\-P\-HPH$_2$P which 
was designed to simulate the ground-state properties 
of the parallel $\beta$ helix of the protein {\em pectate
lyase C}~\cite{yoder,dill3,iba1}. The ground state is known to be low-degenerated.
Up to translations, rotations, and reflections there are only
4 ground-state conformations with energy $E_{\rm min}=-34$.
The density of states ranges over 25 orders of magnitude, and
the ground states were hit frequently with our simulation method such that the
low-temperature properties of this protein could be
investigated with good accuracy. In Fig.~\ref{42C}, 
we show the specific heat and the mean energy of the 42mer.
The specific heat has two peaks, the low-temperature ground-state--globule 
transition occurs near $T_0\approx 0.27$ and the transition between globules 
and random coils at $T_1\approx 0.53$.

In Ref.~\cite{bj1}, we have also compared two 48mers with different
ground-state degeneracies and found also there that 
a pronounced low-temperature peak in the specific heat
only appears for the example with the lower degeneracy
of the ground state (which was about 5000 in that case).
\section{Summary}
We have discussed the relation between the low degeneracy
of the low-lying energy states and the appearance of a 
low-temperature transition
between compact globules and ground states of HP lattice proteins
with 14 and 42 monomers, respectively. For this purpose, we
calculated the density of states of the 14mers by exact enumeration of all 
possible conformations. In order to simulate the density of states
of the 42mer with necessarily high accuracy, we developed
a multicanonical chain growth algorithm that enabled
us to sample the density of states over the entire energy space.
As the main qualitative conclusion we find a correlation between
the degeneracy of low-lying states and the sharpness of the
transition to compact globule states.
\section{Acknowledgements}
This work is partially supported by the German-Israel-Foundation (GIF) under
grant No.\ I-653-181.14/1999 and the EU-Network HPRN-CT-1999-000161
``Discrete Random Geometries: From Solid State Physics to Quantum Gravity''.
\end{document}